\documentclass[letterpaper]{article} 
\usepackage{aaai2026}  
\usepackage{times}  
\usepackage{helvet}  
\usepackage{courier}  
\usepackage[hyphens]{url}  
\usepackage{graphicx} 
\usepackage{natbib}  
\usepackage{caption} 
\usepackage{algorithm}
\usepackage{algorithmic}

\usepackage{newfloat}
\usepackage{listings}
\DeclareCaptionStyle{ruled}{labelfont=normalfont,labelsep=colon,strut=off} 
\floatstyle{ruled}
\newfloat{listing}{tb}{lst}{}
\floatname{listing}{Listing}
\title{Co-constructing sociotechnical AI governance: \\participatory system mapping using algorithm registers}

\author{
    Íñigo de Troya\textsuperscript{\rm 1},
    Maurus Enbergs\textsuperscript{\rm 1},
    Neelke Doorn\textsuperscript{\rm 1},
    Roel Dobbe\textsuperscript{\rm 1}
}
\affiliations{
    \textsuperscript{\rm 1}TU Delft\\

}

\usepackage{outlines}
\usepackage{dingbat} 
\usepackage{booktabs} 
\usepackage{makecell} 

\usepackage{xcolor}
\usepackage{amssymb}

\definecolor{leacolor}{rgb}{0.40,0.60,1.00}
\definecolor{selcolor}{rgb}{0.53,0.30,1.00}
\definecolor{michielcolor}{rgb}{0.9, 0.17, 0.31}
\definecolor{catholijncolor}{rgb}{0.00,0.60,0.30}
\definecolor{maxcolor}{rgb}{0.1, 0.4, 0.8}
\definecolor{ingcolor}{rgb}{0.53,0.10,0.09}

\definecolor{cathacolor}{rgb}{0.670,0.05,0.5}

\definecolor{todonotecolor}{rgb}{0.9, 0.17, 0.31}
\definecolor{changedcolor}{rgb}{0.42,0.27,0.57}
\definecolor{removedcolor}{rgb}{0.867,0.176,0.361}

\newcommand{\nbc}[3]{
		{\colorbox{#3}{\bfseries\sffamily\scriptsize\textcolor{white}{#1}}}
		{\textcolor{#3}{\sf\small$\blacktriangleright$\textit{#2}$\blacktriangleleft$}}
}

\newcommand{\cut}[1]{\nbc{CUT?}{#1}{michielcolor}}
\newcommand{\move}[1]{\nbc{Move?}{#1}{selcolor}}

\newcommand{\wip}[1]{\nbc{WIP}{#1}{catholijncolor}}

\newcommand{\neelke}[1]{\nbc{ND}{#1}{cathacolor}}

\newcommand{\inigo}[1]{\nbc{IMR}{#1}{leacolor}}

\newcommand{\todonote}[1]{\nbc{TODO}{#1}{todonotecolor}}

\usepackage{adjustbox} 
\usepackage{soul} 
\usepackage{svg}
\usepackage{eurosym} 
\usepackage{comment}
\usepackage{enumerate} 
\usepackage{arydshln} 
\nocopyright

\begin{document}

\maketitle

\begin{abstract}
Algorithm registers have been championed as a means of providing transparency on the use of algorithms in public services.
Yet potential publics differ in their expectations of what 
should be made transparent and how, as well as in their interest in and ability to parse the information currently published in the registers.
Moreover, it remains unclear how these instruments can represent the 
sociotechnical systems
in which these algorithms are embedded, and how system-level transparency 
can facilitate accountability.
In this paper, we ask, \textit{what do algorithm registers reveal (and occlude) about the
sociotechnical systems governing algorithmic systems}, and \textit{how can diverse stakeholder perspectives inform a
more pluralistic system-theoretic safety analysis?}
To do this, we probe the municipal algorithm register of 
a Dutch city 
through a case study of 
a decision-support tool for caseworkers' assessment of citizens' welfare benefits eligibility based on legal automation through a business rule engine.
Through 
interviews, surveys, and participatory system mapping workshops (with staff from a municipalility, civil society organisations, and municipal Ombudsmen, N=8), we seek to understand to what extent the register allows stakeholders to map the algorithmic system in question.
These maps
inform a System-Theoretic Process Analysis (STPA) 
that situates 
the register within a wider sociotechnical governance structure.
Participants' contributions 
allow us to identify potential safety hazards which would not have been possible to see using the algorithm register alone,
including benefits eligibility denial, system performance deterioration, and inability to contest wrongful decisions.
By engaging both direct and indirect stakeholders, we reflect on the normative dimensions of algorithm governance efforts and how politics 
shape the practice of system safety analysis.
\end{abstract}



\section{Introduction}


The opacity of algorithmic systems can prevent affected individuals and communities from understanding when they are subjected to erroneous or harmful decisions \cite{janssen2016challenges}, and frustrate efforts by concerned publics to scrutinise these systems for potential risks \cite{burrell2016machine,dingemans2021informatiebehoeften,wieringa2023hey}. 
In the context of algorithmic systems in public services, transparency is a prerequisite for accountability \cite{kroll2017accountable,binns2018algorithmic,nieuwenhuizen2025algorithm}.
High-profile cases like the Dutch childcare benefits scandal (the \emph{Toeslagenaffaire}) have shown that there is a need for more transparency into how these systems function, and for accountability mechanisms that can help restore a sense of justice that is increasingly lost to the digitalisation of the state \cite{Frederik21Correspondent,peeters2023administrative,buszydlik2025understanding}. 
However, what constitutes meaningful transparency remains elusive, particularly concerning efforts to inform the general public \cite{ananny2018seeing}.

In recent years, municipal algorithm registers have become a popular instrument for providing transparency into the use of algorithmic systems in local government \cite{van2024defining,rekenkammer2024kleur,nieuwenhuizen2025algorithm}.
Early adopters
have embraced registers as a platform for informing citizens about how 
algorithms are used 
to provide services such as welfare benefits eligibility \cite{haataja2020public,popa2025frontrunner}.
However, recent empirical studies have shown that there are still open challenges to be resolved before these instruments can 
provide transparency that is meaningful and actionable 
\cite{iaciudadana2025making}.
Both government auditors and academic researchers have found that these registers either disclose too little information to be useful, or are too technical for citizens to 
understand \cite{rekenkammer2024kleur,nieuwenhuizen2025algorithm}.
In Rotterdam, the Court of Auditors concluded that it is simply not clear who the city's register is meant to inform \cite{rekenkammer2024kleur}.
The challenge remains of how algorithm registers can
reconcile the varying information needs of different stakeholders, while respecting organisations’ hesitation to provide full disclosure.
There is a risk that the register may present a further administrative burden on citizens who are ill prepared to parse the information disclosures required to understand how algorithmic systems are used in public services \cite{madsen2022accidental}.
It is clear that more work is needed before algorithm registers provide enough transparency to facilitate accountability \cite{williams2022transparency}.

Despite these shortcomings, algorithm registers can have a ``disciplinary effect" on organisations, serving as a ``meaningful box-ticking exercise" \cite[p. 429]{nieuwenhuizen2025algorithm} that motivates them to take stock of their algorithmic systems, identify their risks, and ensure that appropriate mitigation strategies are in place.
This secondary use suggests that registers may 
help to improve internal awareness about algorithms and their governance, contributing to managing the risks of AI in the public sector \cite{zuiderwijk2021implications}.
To realise this ambition, we require a conceptualisation of algorithms in sociotechnical systems which combines the currently disconnected operational practices and governance instruments that populate the register.
Furthermore, assessing whether strict sociotechnical requirements like safety or fairness are met requires methods to analyze algorithmic tools as embedded in wider sociotechnical systems, to characterize and potentially shape the systemic dynamics across technical and non-technical factors and components~\cite{dobbe2022system,de2025misabstraction}.
Without understanding how algorithm registers are both \textit{a part of} but also \textit{inform} algorithm governance,
the resulting governance practices risk becoming a set of of disparate efforts that cannot meaningfully anticipate or address the actual safety hazards that may emerge in the system.

In this study, we ask two complementary questions. 
\emph{What do algorithm registers reveal (and occlude) about the algorithmic systems and their governance}? (RQ.1.)
Then, \emph{how can diverse stakeholder perspectives 
inform a more pluralistic 
system-theoretic safety analysis?} (RQ.2.)
We answer these questions through a series of workshops, interviews, and surveys with both direct and indirect stakeholders, including civil society organisations (CSOs), Ombudsmen (public mediators), and municipal staff.
In so doing, we seek to understand their expectations, experiences, and desires of algorithmic transparency, and how their unique positionality can help inform a critical safety assessment of a specific decision-support tool listed in the register. The tool selected for this study is called \textit{Avola}, a decision-support tool for caseworkers’ assessment of citizens’ welfare benefits eligibility that is based on legal automation through a business rule engine.
We draw on system-theoretic principles and methods from safety science \cite{leveson2011engineering,leveson2018stpa} to reconstruct the operational and governance architectures in which the decision-support tool and the register are embedded.


By answering these questions we make three core contributions.
First, we provide an empirical assessment of a municipal algorithm register to understand the extent to which it  serves the transparency needs of different relevant stakeholder groups.
Second, we develop and validate a participatory schema for mapping an algorithmic system and its governance structure
by integrating the perspectives, insights and needs of different direct and indirect stakeholders.
Lastly, we reflect on the normative choices and political dimensions behind the abstractions developed while mapping an algorithm governance structure through system safety methods.

\section{Background \& Related Work}

\subsection{The blindspot of algorithm registers: transparency is an essentially contested concept}


While algorithm transparency in the public sector
is thought to
foster citizens' trust in government \cite{grimmelikhuijsen2023explaining}, it remains an essentially contested concept.
Its realisation is highly variable depending on who is disclosing information, and to whom.
In a review of transparency in public administration, Meijer notes that \emph{``transparency may contribute to accountability [...] when there are actors capable of processing the information"} \cite[p.1]{meijer2014transparency}
Transparency is often treated as a performative act -- transparency as something that is done \cite{cellard2020theatres}.
However, it is ultimately a relational act concerning, socially situated actors with differentiated levels of access to information, and ability to understand that information
\cite{murray2025towards}.
Such efforts should thus be evaluated in the context and form in which they are provided -- transparency as something that is given by someone to someone else \cite{felzmann2019transparency}.
Cellard describes algorithmic transparency as a kind of `theatre' play which requires the staging of identities, issues, and algorithms, \textit{``disclosed through the mise en scène of citizens’ motivations, the placing of controversial requests on public bodies, and a regulatory framework redefining administrative procedures as `algorithms'."} \cite[p.1]{cellard2020theatres}
As such, for transparency to meaningfully serve its intended audience, 
algorithmic transparency efforts must grapple with what is to be made transparent, to whom, in what way, and to what ends, 
acknowledging the heterogeneity of the intended publics \cite{axelsson2013public}.

The transparency provided by registers is intended to legitimate the use of algorithmic systems in public services. However, for controversial systems, such as those profiling vulnerable individuals or groups for the distribution of welfare services, such legitimization efforts may nonetheless still be challenged \cite{haug2026algorithmic}.
In discourses surrounding algorithm governance, transparency is an instrumental means towards achieving accountability, such that violations of subjects' rights or freedoms may be identified and addressed in a legitimate public forum \cite{wieringa2020account}.
There is a consensus within the literature
that \emph{``transparency may contribute to accountability [...] when there are actors capable of processing the information"} \cite[p. 1]{meijer2014transparency}
Norval et al. note that if transparency is too highly technical, it may \emph{``undermin[e] the discloure's effectiveness, can disempower subjects, and ultimately hinder broader transparency aims"} \cite[p. 679]{norval2022disclosure}.
They suggest thinking of disclosures as `interfaces', \emph{``designed for the needs, expectations, and requirements of the recipients they serve to inform."} \cite[p. 679]{norval2022disclosure}
However, there is a tension between how much information organisations are willing to disclose and what information different publics need in order to be adequately informed.
The heterogenity of potential publics remains a challenge for public services aiming to provide a one-size-fits-all solution to algorithm transparency \cite{axelsson2013public}.
Without a clear notion of who those actors are, their information needs, and what that transparency should achieve, registers will remain unable to provide meaningful accountability \cite{murray2025towards}.


Critics in both government and academia have proposed shifting the focus away from providing information to citizens and instead towards providing transparency to intermediaries, such as oversight authorities and societal watchdogs \cite{rekenkammer2024kleur,nieuwenhuizen2025algorithm}
- actors more capable of understanding meaningful disclosure and thus of using the register to hold the system providers to account. 
Furthermore, 
these indirect stakeholders may be well positioned to support directly affected parties, by virtue of their proximity, experience, and tact for intermediation \cite{chalke2023implications,dahlvik2022access,madise2024role}. 
Third parties acting on the interest of affected individuals, can serve as a bridge between the lived experience of algorithmic harm and the technical expertise required to unpack it.
However, the challenge remains of what level of information the register should provide in order to serve the information needs of these indirect stakeholders.
Total access is not feasible due to the sensitivity of the data used to make decisions, the fact that source code of underlying algorithms does not explain decisions without said data, and the sheer burden on the deployer organisation to provide all documentation upfront \cite{fenster2005opacity}.
As such, some choices need to be made, and these inevitably lead to the selective inclusion and abstraction of information \cite{paudyal2018algorithmic}.
What information is considered relevant to reveal is thus a political question \cite{de2025misabstraction}.
\subsection{Modeling sociotechnical AI governance as a hierarchical control structure}

The algorithm register serves as a way of ``figuring" the complexity of the Avola system in a way that affected individuals and external stakeholders can understand how the algorithmic system and its governance are structured \cite{andersson2022unpacking}.
The information presented in algorithm registers points toward a whole-system perspective, revealing elements of the technology, its operational procedures, organisational arrangements, and the institutions that govern it. However, many stakeholders still struggle to make sense of this information, meaning that the systems view it suggests remains only partial in practice \cite{rekenkammer2024kleur}. To understand how the various interventions and components relate to one another, a more systematic and integrated approach is needed — one that can reconstruct how these disparate elements fit together. To support this effort, we now turn to system-theoretic methods from the field of system safety.
While initially developed in the context of safety science in industrial applications, such as aviation and energy \cite{leveson2011engineering}, the theories and methods of system safety have recently gained traction in the field of algorithm governance and AI safety \cite{raji2020concrete,dobbe2022system,rismani2023plane,delfos2024integral}.


The System-Theoretic Accident Model and Processes (STAMP) framework models safety and accidents in dynamic complex systems as a problem of inadequate control \cite{leveson2004new}.
Unlike earlier paradigms in safety science, which focused on human error, component failure, or safety culture, STAMP models the emergence of safety hazards and the ensuing accidents as failures to adequately control a system's state in the presence of worst-case environmental conditions \cite{rismani2023plane}.
To do so, STAMP proposes an ontology for modeling safety in sociotechnical systems as a hierarchical control structure composed of control actuators which respond to feedback signals relayed from underlying controlled processes \cite{leveson2011engineering}.
These control and feedback signals span the technical, operational, organisational, and institutional layers of a sociotechnical system, and may include components such as technical risk assessments, human operators with discretionary power, formal recourse channels, and governance instruments such as service-level agreements and regulatory frameworks.
The virtue of STAMP lies in its ability to analyse and orchestrate diverse safety mechanisms across multiple levels through the application of system-theoretic principles \cite{leveson2018stpa}.
STAMP provides the conceptual basis for the System-Theoretic Process Analysis (STPA) method we used in this study, described further in 
the Methods section.

\subsection{Participatory design and mapping}


While the system-theoretic basis of STPA is predicated on the consideration of 
various stakeholder groups, safety science remains a largely expert-led discipline in which the stakeholders who
are brought into the fold
are those already on the inside, capable of contributing their expert knowledge to the safety engineers conducting the assessment \cite{bjornsdottir2023aligning}.
However, 
a growing body of evidence on the harmful consequences of algorithmic systems has shown that affected and indirect stakeholders experience and understand
these systems
in ways that those spared of their outputs simply do not \cite{eubanks2018automating,costanza2020design,katell2020toward}.
The ``participatory turn" in AI \cite{delgado2023participatory} has acknowledged
that 
affected stakeholders 
can bring new forms of situated knowledge that designers may be unaware of \cite{greenbaum1991introduction,haraway1988situated}, enhance public oversight \cite{kallina2025stakeholder}, anticipate risks \cite{kallina2025stakeholder}, reflect on design objectives \cite{katell2020toward}, or help to ensure the systems are useful for those who ultimately have to use them \cite{gutierrez2019explaining,zejnilovic2020algorithmic}. 
Participation is also often seen as a moral imperative 
to give
these systems 
social legitimacy \cite{d2023data,costanza2020design}.

As such, participatory approaches to the design and governance of algorithmic systems can complement safety engineers' own expertise.
Despite this, the role of critical actors outside academia, such as civil society organisations and Ombudsmen 
(public mediators), is underexplored in both the literature on participatory approaches to AI, and in safety science.
While affected individuals and communities may not need technical or legal knowledge to realise that they are experiencing harm or injustice \cite{kuo2023understanding}, they often do need such forms of expertise in order to make their case and seek redress. As such, rather than seek to involve directly affected parties in the design and governance of these systems, we may also consider how to engage other actors which already act as critical governance and safety mechanisms on their behalf \cite{himmelreich2023against}.




\section{Case study: the Avola welfare eligibility algorithm}


De Gemeente (a ficitious name)
is one of the first municipalities in the Netherlands to adopt algorithm registers.
Their municipal algorithm register 
adheres to the goals set for the national algorithm register by the Ministry of Internal Affairs. 
The national algorithm register's website states that
\emph{``The Algorithm Register contains information about algorithms used by the government. This makes this information findable and available to citizens, their advocates, the media and supervisors."} \cite{algoritme_register_nationale}
Among the stated goals of the register are \emph{``Increasing trust in government"} and \emph{"Increasing the controllability of the government"}, proposing that \emph{``If the government shows what it does, citizens and organisations can better control it. The Algorithm Register supports this control by society."}
At the time of writing, 
De Gemeente's
register documents 18 algorithmic systems that are in use, and 2 that have been decommissioned.
The systems are labeled as either high risk or low risk, and range from monitoring improper waste disposal (low risk, decommissioned) to determining eligibility for welfare benefits (high risk, in use).

We probe the algorithm register through a case study of an algorithmic decision support tool that municipal caseworkers use to determine whether or not citizens are eligible for welfare benefits.
The system in question, Avola, is a ``no-code" business rule engine which implements the written laws that govern welfare subsidy eligibility \cite{kaeseberg2019code,escher2024code}.
The resulting law-codification implementation is validated by legal experts.
Frontline caseworkers use Avola as a decision-support tool when they make their own determinations about applicants' welfare eligibility status.
Avola was developed by an independent contractor, Bizzomate (since acquired by CIPHIX; see \cite{blacktrace_mergers}).

The register provides information regarding technical, operational, organisational, and institutional components, describing, among others, the datasets used by Avola to determine benefits eligibility, the role of caseworkers in exercising discretion over final eligibility decisions, information regarding complaints procedures and Freedom of Information Requests 
(Open Government Law, \textit{Wet Open Overheid} [WOO] in Dutch), 
and the legal articles implemented through Avola.
Additionally, it indicates that data protection (DPIA) and fundamental rights (FRAIA) impact assessments have been conducted.
It is worth noting that while the AI Act is set to make the FRAIA mandatory \cite{ecnl2024towards,ecnl2025guide}, it remains a tool to facilitate reflection within the organisation, rather than a strict compliance instrument for external auditors \cite{gerards2022fundamental}.




\section{Methods}


\begin{table*}[hbt!]
  \centering
  
  \fontsize{9pt}{9pt}\selectfont
  \caption{Study design. See Appendix 
  A
  for the interview protocols; Appendix 
  B
  for the worksheet questions.}
  \label{tab:study_design}
  \begin{tabular}{
    p{0.02\textwidth} 
    p{0.11\textwidth} 
    p{0.38\textwidth} 
    p{0.4\textwidth}
    }
    \toprule
    Stage & Description & Purpose & Datasets \\
    
    \midrule

    I & Interviews & Understanding the Algorithm Register \& Avola system & Interviews (audio, transcriptions) \\

    \hdashline[0.5pt/5pt]

    II & Preliminary map & Drafting an initial system map for later Workshop 2 (Stage IV) & Algorithm register entry on Avola, internal documentation (incl. FRAIA), interviews with municipal staff (Stage I), preliminary map $M_{0}$ \\

    \hdashline[0.5pt/5pt]

    III & Workshop 1: Brainstorming & Understanding indirect stakeholders' \textit{expectations} of, \textit{experiences} with, and \textit{suggestions} for the register & Algorithm register entry on Avola, Surveys, Workshop recordings (audio, transcripts), paper annotated with suggestions\\

    \hdashline[0.5pt/5pt]

    IV & Workshop 2: Participatory mapping & Gathering stakeholders' input for later STPA system safety analysis (Stage V) & Workshop recordings (audio, transcripts), worksheet $W$, preliminary map $M_{0}$ (Stage II), annotated maps $M_{0}$* (A1 paper) \\

    \hdashline[0.5pt/5pt]

    V & System safety analysis & Situating the Algorithm Register \& Avola system within the broader sociotechnical system architecture  & Hierarchical control structure, STPA analysis table, annotated maps (Stage IV) \\

  \bottomrule
\end{tabular}
\end{table*}
The study was composed of five stages (Stage I-V), summarised in Table \ref{tab:study_design}.
In this paper we focus on reporting the findings from Stages IV-V.
The contributions made in these later stages were informed by the 
earlier Stages I-III,
which served to familiarise the participants with the different methods and resources used in this study, including the algorithm register and the Avola system,
in order to be able to perform the participatory system mapping in Stage IV.
The study was approved by a Human Research Ethics Council review.

\subsection{Stakeholder identification}

    
    
    



   





In order to capture a plurality of views on Avola and the register, we identified
three relevant stakeholder groups: municipal staff, municipal ombudsman, and civil society organisations.
These groups were chosen by virtue of their expertise on the governance of algorithmic systems in public services, and understanding of their impacts on citizens.
As our primary focus was the algorithm register, not Avola, we did not engage with system developers.

\subsubsection*{Direct stakeholders (Group M)}
As internal stakeholders, municipal staff were those most familiar with both the algorithm register and the Avola system itself.
These included staff in charge of algorithm governance (such as maintaining the algorithm register, conducting risk assessments, etc.) (M1, M2, M4), as well as the product owner (M3), who is in charge of commissioning and managing the development and maintenance of the Avola system, which they see as a means for operationalising municipal policy.
The product owner also provides information and documentation to staff in charge of algorithm governance who publish and maintain the relevant information in the algorithm register.

\subsubsection*{Indirect stakeholders (Group OC)}
Municipal Ombudsmen and CSOs were chosen as external stakeholders who play a vital role in protecting citizens from administrative harms in public services.
Municipal ombudsmen 
mediate disputes between citizens and the municipality, when citizens feel like their concerns are not being appropriately addressed, or struggle to resolve their issues directly with municipal staff (e.g. caseworkers or the complaints department)
\cite{chalke2023implications,dahlvik2022access,madise2024role}.
In the Netherlands, Municipal Ombudsmen are installed by and report back to City Council, who grants them a supervisory mandate over the municipality, 
which is thus compelled to be cooperative during independent investigations.

Civil society organisations 
represent citizens' interests through various forms of policy advocacy work and raising public awareness.
While CSOs have no formal authority over the municipality, they can conduct research to exert influence externally by influencing other actors who do, such as policymakers or ombudsmen.
They may achieve this through policy briefs, public awareness campaign that generate political pressure, and other strategies.

\begin{table*}[hbt!]
  \fontsize{9pt}{9pt}\selectfont
  \caption{Participants' background and involvement in interviews and workshops (see Table \ref{tab:study_design}). The dotted line separates direct and indirect stakeholders (Groups M and OC), who were grouped for the two iterations of Workshop 2.}
  \label{tab:stakeholder_identification}
  \begin{tabular}{
    p{0.04\textwidth} 
    p{0.2\textwidth} 
    p{0.2\textwidth} 
    p{0.1\textwidth}
    p{0.1\textwidth} 
    p{0.1\textwidth} 
    p{0.1\textwidth}
  }
    \toprule
    ID & Organisation & Expertise & Tenure in role (years) & Interviewed (Stage I) & Workshop 1 (Stage III) & Workshop 2 (Stage IV) \\
    
    \midrule
    
    M1 & Municipality & Algorithm governance & 1-5 & \checkmark & \emph{X} & \checkmark \\
    
   M2 & Municipality & Algorithm governance & 0-1 & \checkmark & \emph{X} & \emph{X}\\

    M3 & Municipality & Product owner & 5-10 & \checkmark & \emph{X} & \emph{X} \\

    M4 & Municipality & Algorithm governance & 1-5 & \checkmark & \emph{X} & \checkmark\\

    \hdashline[0.5pt/5pt]
   
    O1 & Municipal Ombudsman & Applied Research & 0-1 & \emph{X} & \checkmark & \checkmark\\

    O2 & Municipal Ombudsman & Applied Research & 1-5 & \emph{X} & \checkmark & \checkmark\\

    C1 & \makecell[lt]{European Center for \\Not-for-Profit Law} & Legal advisor & 1-5 & \emph{X} & \checkmark & \checkmark\\

    C2 & Open State Foundation & Policy advisor & 0-1 & \emph{X} & \checkmark & \checkmark\\
    
  \bottomrule
\end{tabular}
\end{table*}
\subsection{Participatory mapping workshops (Stage IV)}

Two separate mapping workshops were held with each of the groups, Group M and Group OC.
Each workshop lasted 3 hours, and produced multiple sources of data: audio recordings, annotated system maps, and worksheets.
An introductory 20-minute presentation illustrated
an example STPA analysis of a separate welfare benefits eligibility system.
This served to familiarise participants with the STPA method and potential outcomes of the analysis.

Following this, participants were presented with a preliminary system map ($M_{0}$) and a worksheet ($W$).
Map $M_{0}$, drafted by the authors, presented the Avola system components revealed in the algorithm register.
Additionally, some components known to the authors through publicly available documentation were added to the map in order to prompt participants to think about what other components may have been left out of the register.
The map was printed on a large A1 size sheet of paper to allow participants to annotate it freely.
Worksheet $W$ posed 11 guiding questions 
(see Appendix B) 
to help participants to assess the map and contribute to it.
Participants were first given 40 minutes to work individually to complete the worksheet and annotate Map $M_{0}$.
Then,
participants exchanged their insights with others in the group and continued to annotate the map for another 40 minutes.
Group OC had an additional plenary session to share insights among both Ombudsman and CSO staff.
\subsection{System-Theoretic Process Analysis (STPA)}

\label{sec:methods_STPA}

STPA consists of 4 steps, which are performed iteratively, starting at higher levels of abstraction and working towards more nuanced representations of components and their interrelations. 
STPA is typically conducted
in consultation with direct stakeholders such as system designers, managers, policymakers, engineers, and system operators, all of whom have specialised knowledge about different aspects of the system design, deployment, use, and governance \cite{leveson2018stpa}. 
In this study, we propose involving indirect stakeholders due to their commitment to the public interest and their differentiated expert knowledge.
Given the complexity of the STPA, we asked participants to focus on mapping the system (Step 2), and later used their input to inform a full STPA.
The 4 steps are as follows:

\subsubsection*{Step 1: Identify Losses} 
First, the analysts must identify \textit{losses} which are to be prevented.
This requires engaging with direct stakeholders to determine what is at stake in the event of a loss.
For example, a municipality using an algorithmic system to determine welfare benefits eligibility may identify wrongful denial of benefits as the loss to be prevented through the proactive design of safety measures.

\subsubsection*{Step 2: Model the Hierarchical Control Structure}
Next, the safety engineers model the hierarchical control structure (HCS) which describes the sociotechnical governance architecture, which includes both the process within which the algorithmic tool is used and the broader set of processes that contribute to ensuring the system operates within a margin of safety.
In our running example, the municipality may have a series of checks and balances in place to prevent and mitigate wrongful benefits denial, including risk assessments and channels for recourse.
The safety engineers may draw on interviews, technical documentation, or other available means in order to understand and reconstruct the feedback and control mechanisms which keep the system safe during operation.

The HCS is divided into development and operations.
Each block represents an entity which may be a process, instrument, or organisation.
Horizontal arrows represent feedback channels (upwards) and control actions (downwards).
Lateral lines represent information flows.
(For an example, see the final system map in Fig. \ref{fig:map_full}.)

\subsubsection*{Step 3: Identify Unsafe Control Actions}
With the hierarchical control structure in place, the engineers then set out to identify Unsafe Control Actions (UCAs) which are actions that, combined with worst-case environmental conditions, in the presence of systemic hazards, can lead to a loss.\footnote{Different types of UCAs (see \cite[p. 35]{leveson2018stpa}):
1. Not provided when needed;
2. Provided when not needed;
3. Provided at the wrong time (too early or too late);
4. Applied for too long or not long enough;
5. Provided in the wrong order;
6. Provided under the wrong conditions or context;
7. Provided to the wrong component or actuator;
8. Provided with the wrong magnitude or intensity;
9. Provided with incorrect parameters or settings.}
For example, if a human-in-the-loop is given discretionary power to overturn erroneous recommendations provided by a decision-support tool, but fails to do so (perhaps due to automation bias), this would constitute a UCA of Type 1 (``action not provided when needed").
This failure to act, combined with opaque decision-making logic and overtrust in automation (systemic hazards) and a model error (a worst-case environmental condition), would lead to a loss of citizen's benefits eligiblity.

\subsubsection*{Step 4: Identify Loss Scenarios}
Finally, the safety engineers can use the HSC and UCAs to reason through potential scenarios that may play out in which system hazards result in a loss.
For example, a loss scenario may reconstruct how a human-in-the-loop safeguard may break down due to operational pressures.
This requires identifying
system hazards and system constraints to prevent those hazards from emerging.
System hazards are caused by systemic conditions which can contribute to specific hazardous situations.
These can be refined into \emph{sub-hazards} which are localised instances of the broader system hazards \cite[p. 21]{leveson2018stpa}.
For example, the opacity of a decision-making algorithm poses a system hazard because it affects multiple stakeholders.
This manifests as sub-hazards when a caseworker using the algorithm as a decision-support tool, a citizen subject to those decisions, or an ombudsman wants to understand why a decision was made.

\section{Participatory System Mapping}

\begin{figure*}[t!]
\includegraphics[width=500pt]{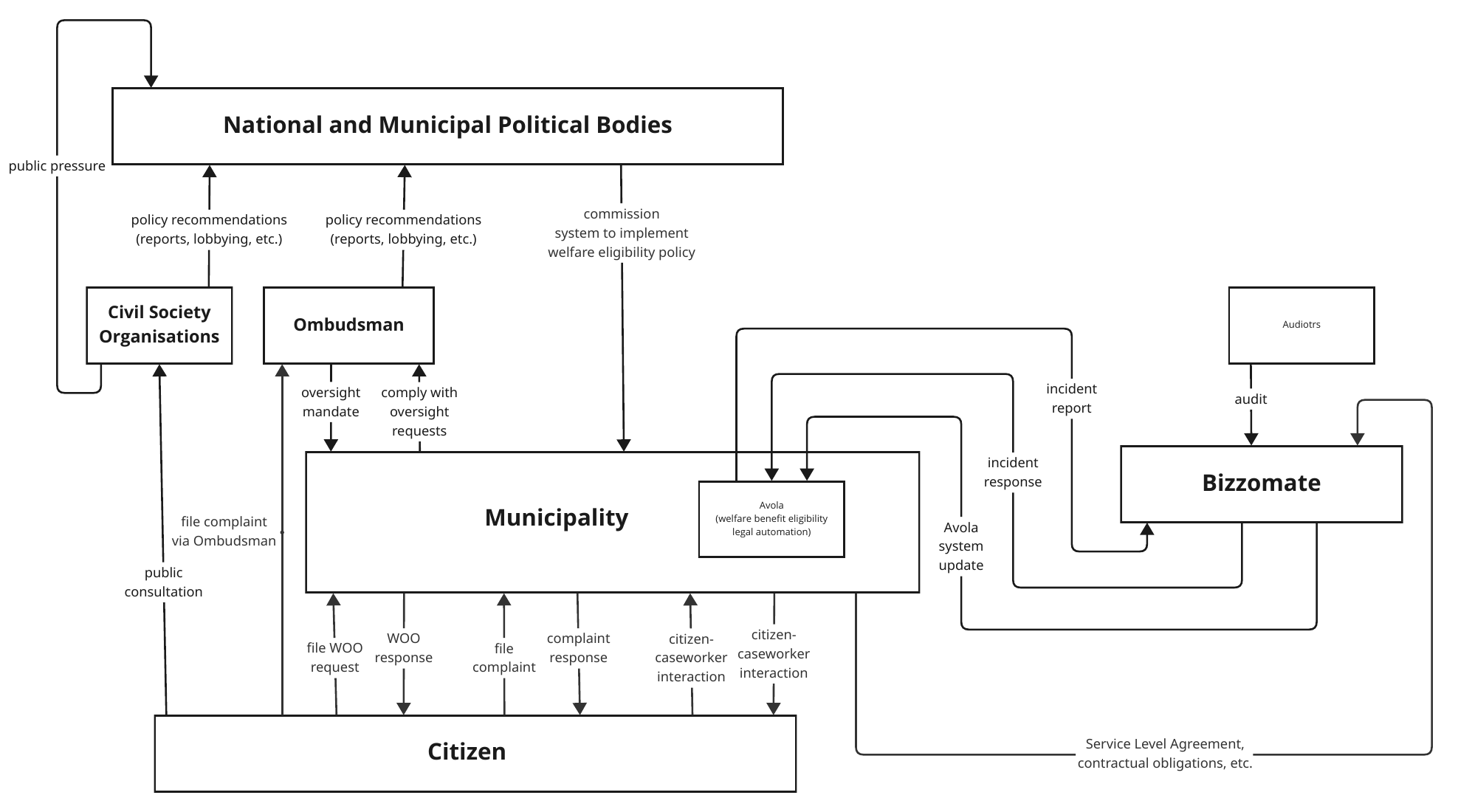}
\caption{
    A high-level representation of the hierarchical control structure of Avola, showing the control (downward arrows) and feedback (upward arrows) relationships between different stakeholders. (Please see the Supplementary Material for a larger version of the map, and a detail of the hierarchical control structure within the municipality.)
}
\label{fig:map_full}
\end{figure*}



The participatory mapping workshops resulted in six system maps, one by each participant.
These 
were combined into one by the authors, as shown in Fig. \ref{fig:map_full}.
In so doing, the final map combined the plurality of participants' subjective and socially situated understandings of the Avola system and its governance architecture into a single multifaceted representation that illustrated the diversity of their concerns and expertise.
Participants' contributions ranged from adding missing actors, organisations, institutions, and links among these, to points of interrogation on parts of the map which were not explained by the algorithm register alone.
The contributions made by each stakeholder group were informed by their positionality -- their role in relation to the algorithmic system, its governance structure, and the citizen.
This influenced what aspects of the map they thought were missing, and what areas of the map they raised further questions on. 
The colour-coding in Fig. \ref{fig:map_full} shows clearly how distinctive these contributions were among the groups.
Below, we summarise these distinctions.





\subsubsection*{Civil Society Organisations}
Drawing on their experience working on civil advocacy regarding social impacts of algorithmic systems and digital rights,
CSO staff's contributions and questions focused on technical aspects of the system, algorithm governance instruments, as well as complaints channels.
These included expanding the range of complaints channels beyond the municipal complaint channel mentioned in the register to also include those available through the Dutch Data Protection Authority (Autoriteiten Persoonsgegevens), the AI Act, Civil Court, and the Ombudsman.
They pointed out that the Freedom of Information Request (WOO) would likely prove a high barrier to access for citizens, and was more likely to be used by expert users such as themselves, Ombudsmen, journalists, or academics.
This remark showed how the register already provides some information which is intended for more expert publics rather than for citizens subject to algorithms like Avola.

Regarding the risk assessments (both DPIA and FRAIA), CSO staff questioned who had been involved in conducting them, or consulted during the process, and whether these assessments informed each other. They also asked whether these had been conducted once or multiple times, and noted that it was not clear when such assessments had taken place, or whether they were still relevant for the given software version of Avola currently in deployment.
Regarding technical aspects, CSO staff asked about what the Avola incident reports contained (whether they were merely error logs or caseworkers' assessments of individual citizens' cases), and what was done in response to these reports.
They also pointed out that the little information published in the register about what Avola did said very little about how it was actually implemented in practice (i.e. what type of algorithm it was, what features it used, etc.)
Lastly, CSO staff questioned what accountability body Bizzomate was subject to.

\subsubsection*{Municipal staff}
The contributions by municipal staff focused on the role of various political bodies in governing the system, as well as introducing nuance that was not available through the algorithm register.
For example, M1 and M4 explained that the Aldermen (members of the municipal legislative body) were in charge of commissioning both the Avola system itself, as well as the algorithm register.
The algorithm register team and Avola's product owner report back to the Aldermen such things as the results of the risk assessments, which the Aldermen must then rely on to determine whether to approve or reject these systems prior to deployment.
The Aldermen in turn report to the City Council, above which sit the Ministry of Interior, 
the House of Representatives, and finally the Senate.
For matters concerning the municipal algorithm register, CSOs and Ombudsmen may engage with the Aldermen or City Council directly.
However, 
should they identify deeper issues stemming from the national register, CSOs and Ombudsmen may address the political bodies higher up in the hierarchy.
Additionally, any public actor 
may choose to seek redress through Civil Court against either the municipality or the third-party developers of the Avola system itself.



\subsubsection*{Ombudsman staff}
Ombudsman staff added the least to the maps, likely due to their reservations about how well mapping served their own operational needs and their perspective of how they seek to understand the systems they examine.
O2 noted that \textit{``our work is more about the human interest and this is more the systems [point of view]. A completely different way of thinking."}
O1 illustrated this further, noting that the citizen-caseworker interaction was \textit{``the only human connection"} on the map.
Further, O2 stated that while the algorithm register gives \textit{``the illusion of explanatory power"}, Ombudsmen are more interested in understanding how the citizen is affected than how the system works.

The components they did contribute consisted of support organisations (focused on youth), as well as counsellors who citizens could seek help from. 
Beyond these few additions, O1 and O2 mostly used map as a prompt to ask questions, such as, \textit{is it clear for the citizen which data are being used?} (O1)
Additionally, they questioned the sufficiency of some of the register's disclosures, noting that it was not clear what department handled the Freedom of Information Request, thus making them doubt how effective such a channel could be (O1).
This enabled us to consider what frustrations citizens may feel when using the register themselves.

\section{System safety analysis}
\label{sec:results_stpa}

Building on the contributions from the participatory mapping workshops, we next conducted the full STPA system safety analysis.
Here, we share an overview of the STPA in a narrative form that illustrates how successively implicating different system actors and components at increasing levels of the hierarchy allows us to gradually broaden the sociotechnical frame -- across the technical, operational, organisational, and institutional frames.
We do so through a series of \textit{Loss Scenarios} which span (A) the human-in-the-loop, 
(B) the complaints procedure, 
(C) the Ombudsman's oversight mandate, 
and (D) political pressure and the public sphere.
We also note where participants' contributions in the precending stages (I-IV) informed these scenarios.

In Table \ref{tab:stpa_excerpt}, we provide an excerpt of the full STPA which breaks down the Loss Scenarios into the corresponding Losses, System Hazards, Subhazards, and UCAs.
These are indexed for later reference in the text (L-1, etc.).

\begin{table*}
   \centering
  \fontsize{9pt}{9pt}\selectfont
  \caption{Excerpt of the STPA analysis.}
  \label{tab:stpa_excerpt}
  \begin{tabular}{
    p{0.12\textwidth} 
    p{0.12\textwidth} 
    p{0.26\textwidth}
    p{0.24\textwidth}
    p{0.15\textwidth}
  }
    \toprule
    Loss & 
    System Hazard & 
    Subhazard & 
    Unsafe Control Action (UCA) & 
    Loss Scenario\\
    
    \midrule

   L-1: Denial of benefits eligibility &
SH-1.1 Opaque decision logic & 
    SH-1.1.b. Automation bias leads to overreliance on Avola by caseworker &
    U-1.1.b.i. Caseworker approves erroneous Avola recommendation &
    A. Human-in-the-loop\\

    \hdashline[0.5pt/5pt]

    L-2: System performance deterioration &
    SH-2.1: Failure to identify errors & 
    SH-2.1.a. Failure to learn from aggregated complaints &
    U-2.1.a.i. Change request not triggered U-2.1.a.ii. Designer fails to correct model & 
    B. Complaints procedure \\

    \hdashline[0.5pt/5pt]

    &
    & 
    SH-9.1.a. FRAIA does not report error metrics and performance &
    U-9.1.a.i. CSOs unable to contest Avola on technical grounds & 
    D. Public sphere \\

    \hdashline[0.5pt/5pt]

    L-3: Inability to contest decision &
    SH-3.2. Lack of sufficient information (e.g. in Register) & 
    SH-3.2.b. Register lacks a well-defined audience &
    U-3.2.b.i. Citizen unable to contest Avola via complaints procedure &
    B. Complaints procedure\\

    \hdashline[0.5pt/5pt]

    &
    & 
    SH-3.2.c. Register lacks supporting documentation (e.g. FRAIA and other risk assessments, technical info including error rates, etc.) &
    U-3.2.c. Ombudsman unable to respond to citizen complaint &
    C. Ombudsman oversight mandate\\

  \bottomrule
\end{tabular}
\end{table*}
\subsection{Loss Scenario A: the human-in-the-loop}

The algorithm register insists that Avola does not make any decisions automatically, but rather serves as a decision-support tool for caseworkers 
that act as a human-in-the-loop who has the final say.
However, as much literature has shown (e.g. \cite{green2019principles,ruschemeier2024automation}), this reliance on the human agent to steer the algorithm rests on strong assumptions (e.g. that the caseworker will be able to catch the algorithm's errors, for example, by understanding what factors influenced its output) and are prone to error in many ways (e.g. caseworkers may feel compelled to accept algorithmic outputs perceived as value-neutral or objective).
CSO staff C1 noted that while the register stressed the role of the caseworker as the ultimate decision-maker, it could not portray the conditions under which caseworkers made those decisions (perhaps pressured by time constraints to not disagree with Avola and having to write a justification report under vigilance from the Quality Assurance staff who check whether deviations from Avola's recommendation are justified).
Despite this limitation, which may prove beyond the scope of the register's capacity, we may interpret the assertion of the caseworkers' discretion as a comforting but shaky reassurance, hiding from citizens such issues as algorithmic bias, opacity, and organisational pressures, which may betray the fact that Avola's innocuous function as a decision-support tool may be superseded by the perceived objectivity of its logic, whether in the eyes of caseworkers themselves or of superior managerial staff from whom they receive this narrative 
of automated administrative neutrality.

We may consider a hypothetical situation in which 
citizen data which is out-of-date 
(see map: ``[BRP/Suwinet check fails due to out-of-date data]", raised as a potential issue by Ombudsman staff O1),
which may occur for various reasons, including human error (e.g. citizens failing to update their own data) or technical issues (e.g. internal server errors pertaining to database maintenance).
Automation bias (SH-1.1.b.) may lead caseworkers to over-rely on Avola, for example, because they see it as a legal automation model that simply ``implements the law" and is thus not subject to spurious data biases or other issues that beset data-driven algorithms.
Whatever the case may be, it falls upon the caseworker to recognise and identify these errors as such.
Should they fail to do so (U-1.1.b.i.), the human-in-the-loop safeguard falls apart, and the citizen may be wrongfully denied benefits for which they are, in fact, eligible (L-1).
In this scenario, a technical error (BRP/Suwinet check fails) leads to an operational error (casework misses a technical error), which is a result of an organisational safety culture of undue 
trust in legal automation as error-free and objective.




\subsection{Loss Scenario B: the complaints procedure}

As a result of Loss Scenario A, a citizen may seek recourse for loss L-1 by filing a complaint through the municipality.
Complaints are triaged among the relevant departments and then entrusted to the caseworker responsible for the case (as per our interview with G1).
However, CSO staff C1 noted that it wasn't clear from the register how complaints would be handled in practice, leaving them in doubt as to whether filing complaints would 
be an efficient form of recourse.

In the FRAIA documentation, we found that Avola shares incident reports with Bizzomate.
However, CSO staff C2 noted that it was not clear what these error reports entailed (e.g., whether they relate to the individual case or are simply technical logs).
Furthermore, the FRAIA documentation noted that error metrics were not tracked.
As such, it is possible that complaints are treated on a case-by-case basis, and not aggregated and analysed in a way that would reveal 
systematic error patterns.
Failing to identify these systematic errors would mean that the appropriate change requests are not filed (U-2.1.a.i), and thus developers would fail to correct for the errors and deploy a model update (U-2.1.a.ii).






\subsection{Loss Scenario C: the Ombudsman's oversight mandate}

If the citizen is not satisfied with how their complaint was handled, they may choose to take their issue to the Ombudsman, who can act as a mediator between the citizen and the municipality.
One clarification which Ombudsmen may wish to pursue is how the implementation of legal rule automation has been implemented in Avola.
In response, municipal staff may provide the FRAIA documentation which mentions that \textit{``legal tests"} are performed
prior to deployment.
However, the FRAIA does not provide any detail as to what those tests entail, preventing the Ombudsman from using existing documentation to scrutinise the implementation details of Avola.
As a result, the Ombudsman may be unable to respond appropriately to citizens' complaints (U-3.2.b.i.).



\subsection{Loss Scenario D: political pressure and the public sphere}
Eventually, CSOs and Ombudsmen may also choose to pursue legal action through Civil Court (suggested as a potential course of action by G4).
However, in order to mount a legal case (e.g. against the developer of Avola), they would need to supply evidence of wrongdoing.
Both CSO and Ombudsman staff pointed out that the register's explication of the FRAIA and DPIA assessments was altogether insufficient.
The register simply stated that these had been conducted ("FRAIA: Yes; DPIA: Yes") but withheld any further elaboration as to who was involved and what they found 
While one may assume the risk assessments had not yielded any problematic outcomes, even this is left to the imagination.

Thanks to the cooperation of municipal staff, we had the privilege of consulting the FRAIA documentation ourselves.
While we did not find any evidence for fundamental rights violations, some of the answers provided to questions on the validation and evaluation of Avola's performance left much to be desired.
For example, when asked about what error metrics were used to evaluate Avola (FRAIA question 2B.3.6.), it was simply stated that, \textit{``it happens, but it is so marginal that no number can be attached to it."} 
This lack of detail is likely due to insufficient expertise or resources to conduct the FRAIA (as per our interview with M4).
(While Bizzomate may have a rigorous evaluation procedure in place, this is simply not properly reported in the FRAIA where prompted to do so.)

This lack of adequate technical documentation regarding the performance of Avola means that, even after filing a Freedom of Information Request, external independent observers, such as Ombudsmen or CSOs, may be unable to provide evidence of wrongdoing in order to bring forth a credible legal case in Civil Court (U-9.1.a.i.).

\section{Discussion}



\subsection{On the role for public actors in realising the ambitions of the algorithm register}






Amid increasing debate about enabling public oversight of algorithmic systems \cite{Sloane2020,himmelreich2023against,williams2022transparency}, we find that the algorithm register falls short of these ambitions in practice.
For both actors focused on the impacts for citizens, like Ombudsmen, and those concerned with the innerworkings and governance of the algorithms, like CSOs, the algorithm register raised more questions than it answered about Avola and its governance.
While the stated goal of the algorithm register is to inform the public about how algorithmic systems are used and governed \cite{algoritme_register_nationale}, we found that the information they reveal is too abstract to be meaningful for external actors.
The declaration of risk assessments such as FRAIA or DPIA is underwhelming to actors who are interested in understanding what those assessments actually found.
Furthermore, while the FRAIA instrument is meant to catalyse reflective discussions among a diverse range of stakeholders, including civil society organisations and citizens, we found that, in this case, the municipality did not have the capacity to facilitate such forms of participation.
Once we gained access to the FRAIA documentation, we found it to be limited for certain critical points, like evaluation.
As such, the mere declaration of risk assessments may give the public a false sense of security.
While these documents can be obtained through a Freedom of Information Request (WOO), having to do so provides an obstacle to accessibility, in an instrument for which the intended purpose is to inform the public.

Nevertheless, these instruments are an important step in making algorithms and their governance more accountable to the public.
As we have shown in this study, scrutiny from indirect stakeholders, such as Ombudsmen, civil society organisations, and academic researchers, can help to identify the affordances and limitations of these efforts.
The openness of the municipality to participate in this study bears testament to their 
willingness
to reflect and improve on their algorithm governance practices.
We recognise that the duty of documentation (e.g. through the algorithm register or the FRAIA) can be a burden on already under-resourced public organisations.
As a result, it is understandable that these efforts may fall short of the expectations of more critical actors.
While most public engagement efforts concerning algorithm registers' design requirements have sought to involve the general public \cite{dingemans2021informatiebehoeften,ICTU}, we encourage greater inclusion of expert communities such as civil society organisations, Ombudsmen, independent researchers, journalists and other actors with the capacity to make use of the information 
and to grapple with the heterogeneity of affected publics \cite{axelsson2013public}.
At the same time, we caution 
external actors to be mindful of public organisations' resource challenges, and avoid diminishing such efforts as disingenuous \cite{ananny2018seeing,cath2021dutch}.




\subsection{System mapping as a participatory practice}


In our attempts to reconstruct the governance architecture of the Avola system, we found that the different stakeholder groups' input significantly enriched the information available in the algorithm register.
While the register is intended to inform the public about what safeguards are in place to mitigate potential risks 
(e.g. the FRAIA and DPIA assure the public that potential infringements of data protection and fundamental rights have been assessed and safeguarded against), the potential Loss Scenarios that we found would not have been evident relying on the algorithm register alone.
The practice of sharing our process with participants enabled us to perform a more comprehensive safety analysis of the system, and to appreciate the forms of information that the register would need to divulge in order to truly function as an instrument that can support public accountability and control.

While the initial goal of the mapping exercise was to inform the STPA safety analysis, we found that the practice of mapping itself can also be seen as a valuable intervention.
Mapping allowed participants to ask questions to each other, and to the experts in the room.
These questions in turn helped the safety experts to review their assumptions, for example, about the value of mapping for different participants' existing practices.
As such, we encourage safety experts to view mapping to as a means to an ends, but as an ends in itself -- to see mapping as an opportunity to invite critical reflection that can inform their own expert safety analysis, and to empower public actors to understand and scrutinise algorithmic systems and their governance.





System-theoretic methods from safety science, such as STPA, can help indirect stakeholders to make a compelling case for why the inclusion of certain contextual factors is significant for informing the public about how these risks are mitigated against.
At the same time, while
these methods
provide an ontological basis through which to describe the sociotechnical context in which such algorithms are deployed, these methods also impose a certain understanding of these systems onto the world.
As both Ombudsman staff noted, system maps lack the ability to tell the human story, and as such, they are unlikely to speak to the concerns of affected citizens.

\subsection{Beyond algorithm registers: mapping sociotechnical systems and their politics}




Beyond documenting algorithmic systems as technical artifacts, algorithm registers also detail the operational, organisational and institutional processes surrounding their use and governance.
In so doing, they bear witness to the complex sociotechnical systems in which algorithms are embedded, and which are required to make them function \cite{chen2024explainer}.
As shown in our system safety analysis, safety is not something that can be achieved by technical means alone \cite{nouws2023diagnosing,dobbe2025ai}.
Rather, safety requires the orchestration and cooperation of multiple actors, procedures, and organisations \cite{de2025misabstraction}.
As such, algorithm registers play an important role in illustrating how algorithmic systems can be better understood as sociotechnical systems, requiring technical and social components to be jointly designed \cite{baxter2011socio,chen2024explainer}.
The documentation of measures such as channels for filing complaints, discretionary human oversight, and an adequate legal basis serve to contextualise the development and operation of the algorithmic system within the wider scope of governance instruments and resources already existing at the municipality \cite{dourish2004we,de2025misabstraction}.


At the same time, what components are accounted for, and how, is subject to political forces among socially situated actors \cite{selbst2019fairness,de2025misabstraction,mahroof2025navigating}.
Every actor has knowledge about different aspects of the system:
municipal staff were able to elucidate the political structures governing the municipality and its efforts, Ombudsmen were able to speak to the concerns of citizens, and CSOs' critical eye revealed many shortcomings in how the register represented the governance structure around Avola.
However, the municipality is ultimately subject to political actors, such as Aldermen and the City Council, who depend on expert counsel to inform their decisions (including through policy advocacy by Ombudsmen and CSOs).
As such, the register is bound to reflect the public concerns prioritised by these political actors.
External or indirect stakeholders
who wish to expand the scope or detail covered in the register must then seek to advocate for these changes through appropriate channels such as by policy advocacy efforts.



\section{Conclusion}

Algorithm registers are an important step in facilitating transparency on the use of algorithms in public services.
As public organisations continue improving these instruments, there is an opportunity for diverse stakeholders to engage with the institutional design of algorithmic governance.
While algorithm registers have been criticised for lacking a clear audience, and not having teeth, it is important to recognise the unresolved challenges that need to be overcome in order to make these instruments more useful.

In order to be effective tools for accountability, algorithm registers need to be better aligned with the information needs of different publics.
System-theoretic methods like STPA can help to show how algorithm registers integrate with existing governance practices, and can be used to guide a participatory critical reflection that meaningfully engages vital public actors, like CSOs and Ombudsmen.
Our analysis highlights that we still lack sufficiently developed conceptual frameworks for thinking about, designing, and governing sociotechnical systems in an integrated manner. 

We have also demonstrated that mapping and documentation practices are inherently political. What becomes visible in a register reflects negotiations among actors with different roles, forms of expertise, and institutional power. This raises a final, critical question: who has a say in what gets documented?
At the same time, our study of the algorithm register insists that
algorithmic systems must be understood across the technical, operational, organisational, and institutional contexts that make up the sociotechnical systems in which they are embedded.

\appendix
\newpage
\clearpage

\section{Appendix A: Interview protocols}

\label{sec:appendix_interview_protocols}

 \subsection{Interview protocol regarding the algorithm register}

\noindent
\fbox{%
    \parbox{\textwidth}{%
        \textbf{Purpose of the register}
        
        Q1. What is your role at the municipality?
        
        Q2. What is the intended purpose of the municipal algorithm register?
        
        Q3. Are the objectives of the municipal register the same as those of the national register?

        \textbf{Reflecting on the current register}
        
        Q4. How and to what extent are the objectives stated in the national register being met by the municipal register?
        
        Q5. What do you think the register currently does well?

        \textbf{Desired future state of the register and challenges to get there}
        
        Q6. What do you think can be improved?
        
        Q7. What are you currently doing to improve the register?
        
        Q8. What challenges have you faced in implementing those improvements?
        
        Q9. Do you have any further comments that you believe I may have overlooked in my questions?
    }
 }

 \subsection{Interview protocol regarding the Fundamental Rights Impact Assessment (FRAIA)}

\noindent\fbox{%
    \parbox{\textwidth}{%
        Q1. Can you please introduce yourself and say a little bit about your role at the municipality?
        
        Q2. Who else was involved in the FRAIA of the Avola system, besides yourself?
        
        Q3. What insights did the FRAIA process give you about the system? (e.g. lack of change management procedure; \\"instruments for monitoring, steering and accountability are missing or still under development"; etc.)
        
        Q4. Were any actions taken after doing the FRAIA? 
        
        Q5. What do you do if you are unable to answer a question?
        
        Q6. Once the FRAIA is first completed, is it reviewed by someone else before it is declared finished?

        Q7. How can citizens "request their records that include advice from Avola"? (3.6.2) How are they made aware that they can do this?

        Q8. Did you raise any issues found through FRAIA with Bizzomate? How did they respond?

        Q9. Why was it decided to deploy the system despite there not being "proper tools provided for evaluation, auditing, and assurance of the algorithm"? (3.7.1.)
    }%
}

\newpage
\clearpage

\section{Appendix B: Worksheet}

\label{sec:appendix_worksheet}

\noindent\fbox{%
    \parbox{\textwidth}{%
        
        \textit{Q1.} Where would you place yourself on the map?	
        \\ \\
        \textit{Q2.} What components of the map are you connected to, and in what way?	
        \\ \\
        \textit{Q3.} What is missing from the map? (e.g. components, connections)	
        \\ \\
        \textit{Q4.} What other elements of the map come into view through your presence on the map? 
        
        (e.g. other organisations, mechanisms, procedures, etc. that you are ore directly connected to)	
        \\ \\
        \textit{Q5.} What part of the map would you like more detail on? If so, why?	
        \\ \\
        \textit{Q6.} What abstractions or assumptions did you have to make?	
        \\ \\
        \textit{Q7.} In your expereince of working with complaints procedures, is there any information missing from its representation on the map?	
        \\ \\
        \textit{Q8.} What other experiences did you draw on when mapping? 
        
        (e.g. working with a citizen on a similar case in the past...)	
        \\ \\
        \textit{Q9.} What additional resources would you use to map the system (besides the register)?	
        \\ \\
        \textit{Q10.i.} Are there any limitations to the mapping approach we used which make it hard to represent certain aspects of the system?	
        \\
        \textit{Q10.ii.} If so, do you have any suggestions of what can be improved and how?	
        \\ \\
        \textit{Q11.} What value do you see in mapping of algorithmic systems for understanding their governance structure and potential negative impacts? In other words, how could a map inform your work? (e.g. identifying points of intervention, lack of detail about specific areas; helping me to see the bigger picture so that I know where to focus)
    }%
}

\newpage
\clearpage

\section{Appendix C: Hierarchical Control Structure}

\begin{figure}[b!]
\includegraphics[width=520pt]{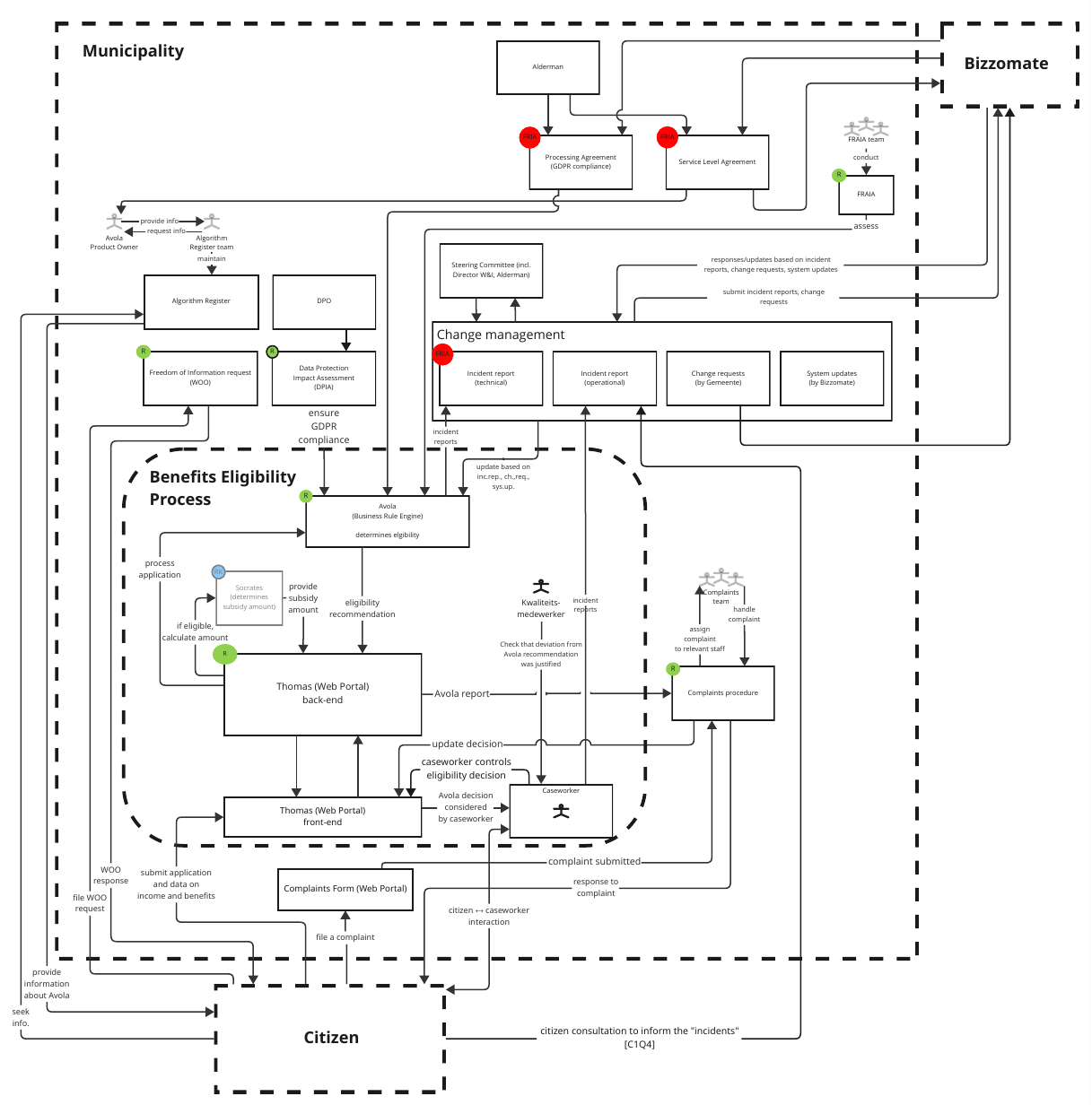}
\caption{
    A detail of the hierarchical control structure of the municipality, part of the design and governance structure around the Avola system.)
}
\label{fig:map_full}
\end{figure}

\end{document}